# Interface chemical and electronic properties of LaAlO$_3$/SrVO$_3$ heterostructures

Arnaud Fouchet*[1,†], Julien E. Rault[2], Mickaël Allain[1], Bruno Bérini[1], J.-P. Rueff[2], Yves Dumont[1], Niels Keller[1]

[1]*Université Versailles St Quentin, CNRS, Groupe d'Étude de la Matière Condensée (GEMaC), 78035 Versailles, France.*

[2]*Synchrotron-SOLEIL, BP 48, Saint-Aubin, F91192 Gif sur Yvette CEDEX, France.*

*Corresponding Author:* *E-mail:* **arnaud.fouchet@ensicaen.fr**

*Present Addresses:*

[†] *Laboratoire de Cristallographie et Sciences des Matériaux (UMR 6508), Normandie Université, ENSICAEN (Ecole Nationale Supérieure d'Ingénieurs de Caen), UNICAEN (Université de Caen), CNRS, 6 Bd. Maréchal Juin, F–14050 Caen, France.*

**Abstract**

We have studied the chemical and electronic properties of LaAlO$_3$/SrVO$_3$ ultrathin films by combining hard x-ray photoemission spectroscopy and transport measurements. We compare single SrVO$_3$ (SVO) ultrathin films and SrVO$_3$ buried below a polar LaAlO$_3$ (LAO) thin layer, both epitaxially grown on SrTiO$_3$. While ultrathin films (4 unit cells) of SVO do show insulating behavior over the entire temperature range, the LAO/SVO interface has a resistivity minimum at 250 K. When increasing the SVO layer thickness, the minimum is observed to shift to higher temperatures, but the resistivity stays always smaller than that of comparable SVO single films. Hard x-ray photoemission spectroscopy reveals a surface or interface related V$^{5+}$ component in the V *2p* spectra for SVO films and LAO/SVO heterostructures, respectively, attributed to a strongly oxidized component. This chemical reconstruction is weaker in LAO/SVO heterostructures compared to single SVO films. We show that this dead layer in SVO ultrathin films has to be considered when the film thickness reaches the few unit-cells limit and propose solutions on how to prevent this detrimental effect.

# 1) Introduction

With the improvement of thin films growth, novel properties at oxide interfaces appear and lead to the development of new functionalities through material confinement and engineering[1].

Understanding these properties brings up new challenges including growth control and film characterization. Unexpected chemical and physical phenomena may appear in oxide ultrathin films where electronic correlations, structure, strain, intermixing, oxygen deficiency, surface reconstructions, and other effects may influence the properties of the heterostructures.

Among these properties, the Mott metal-insulator transition (MIT) is one of the most fascinating phenomena in condensed-matter physics. It has been widely studied for perovskite structures where the MIT can be controlled by cationic substitution, and more recently by dimensional cross-over. According to the Mott-Hubbard theory, the MIT is controlled by the ratio U/W where U is the on-site Coulomb repulsion and W is the electronic bandwidth. In this context, $SrVO_3$ (SVO) is a prototypical system to study the MIT, because of his simple electronic configuration ($3d^1$). Chemical substitution in $Ca_xSr_{1-x}VO_3$ of Sr by Ca has already been used for Bandwidth-controlled MIT (BC-MIT), controlling the strength of the electronic correlations *U/W* without changing the band filling[2,3]. More recently, calculations have predicted that control of dimensionality induces also a MIT with novel properties such as Mott insulator ferromagnetism[4]. The effective bandwidth is decreased by dimensional cross-over from 3D to 2D in ultrathin films related to the reduction of the effective coordination number at interfaces or surfaces.

The presence of a MIT in ultra-thin films of SVO has been confirmed by Angle Resolved PhotoEmission Spectroscopy (ARPES)[5]. For film thicknesses above 6 unit cell (≈ 2.3 nm), the SVO valence band shows metallic states. By progressively decreasing the SVO thickness, a pseudo-gap is observed at the Fermi energy ($E_F$) through weight transfer from the coherent part to the incoherent part of the spectral function. The pseudo-gap finally evolves into an energy gap for two units cell indicating MIT in $SrVO_3$ ultrathin films[5]. By using transport measurements, our group has shown the pseudo-gap appearing for SVO films with thicknesses from 7 nm (18 unit cells u.c.) to 1.6 nm (4 u.c.). In this transition region, transport properties evolve from metallic to localized behavior, where the upturn of the resistivity can be explained by quantum corrections of the conductivity and the variable range hopping model. Furthermore, spin-orbit coupling appears at low thicknesses due to the confinement[6].

Gu et al.[7] reported also a crossover thickness around 7 nm and the presence of a pseudo-gap for SVO thin films deposited on $(LaAlO_3)0:3(Sr_2AlTaO_6)0:7$ (LSAT) by pulsed electron beam deposition (PED). This pseudo-gap was attributed to local metallic states in a globally insulating region.

Interestingly, there is a significant discrepancy between the thickness limit of the MIT studied by ARPES and by transport measurement of a factor 2.

Modification of electronic properties is usually induced by heteroepitaxy influencing the crystalline structure of the film. Indeed, lattice mismatch is changing the crystalline symmetry and bond lengths. Nevertheless, the same substrate ($SrTiO_3$) has been used for our transport measurements and ARPES measurement by Yoshimatsu et al.[5]. Furthermore, by comparison of transport measurements of SVO films grown on LSAT (tensile strain of 0.7%) [7] and STO (1.6%) [6], the same thickness dependence was obtained. Moreover, in superlattices[8,9] and heterostructures[10] the MIT seems also to happen around 2-3 u.c.. Since heteroepitaxy does not seem to be the main driving force behind the differences observed concerning the MIT, chemical disorder should also be taken into account for vanadate films.

In the case of $CaVO_3$ and SVO single crystals, a strong enhancement of *U/W* in the surface region of the single-crystal compared to the bulk has been observed. This enhancement was attributed to two possible origins: reduced atomic coordination at the surface and/or surface reconstruction, which would give rise to a decrease in W and an increase in U compared to the bulk [3].

Thus, the investigation of the surface properties of ultrathin films is crucial for the understanding of the electronic properties of ultrathin film and the thickness limit of the MIT. Even more interesting is the effect of confining the SVO film between the $SrTiO_3$ substrate and a wide band-gap insulator such as $LaAlO_3$ to understand the role of polar interfaces on the SVO properties.

Hard X-ray photoemission spectroscopy (HAXPES), in which the photon energy is tuned into the multi-keV regime, achieves greater sensitivity to study buried layers and interfaces[11,12]. This technique is non-destructive and available in an increasing number of synchrotron radiation facilities. Being furthermore compatible with operating conditions such as electric field bias and ambient pressure [13–15], it is a very attractive technique for studying functional interfaces. For instance, Suga *et al.* used HAXPES on the close-related compounds $VO_2$ and $V_6O_{13}$ to reveal a surface layer with higher resistivity compared to bulk [16,17]. Mossanek *et al.* combined HAXPES and cluster model calculations to follow the spectral weight transfer from coherent to incoherent features for the single crystals of the $SrVO_3$-$CaVO_3$-$LaVO_3$-$YVO_3$ series, pointing out the limitation of the Mott-Hubbard theories[18]. However, there is no hard X-ray photoemission spectroscopy investigation of SVO thin films and LAO/SVO heterostructures to the best of our knowledge.

In this paper, we compare the thickness-dependent transport properties of $SrVO_3$ single films and $LaAlO_3$/$SrVO_3$ heterostructures grown by Pulsed Laser Deposition (PLD). We measure a strong thickness dependence of the MIT in both cases, but the MIT persists down to two SVO unit cells in the LAO/SVO heterostructure while it disappears below 4 u.c. for single SVO films. By using hard X-ray photoemission spectroscopy, we directly probe the chemical properties of the SVO surface in the single films and the SVO/LAO interface in the heterostructures, showing that transport is altered by a electronic and chemical reconstruction in these areas. This reconstruction is less strong at the LAO/SVO polar interface, explaining the lower thickness limit of the MIT.

## 2) Experiments

Due to the good lattice matching between bulk perovskites $SrTiO_3$ (STO) (3.905 Å), $SrVO_3$ (3.84 Å) and pseudocubic $LaAlO_3$ (3.78 Å), heterostructures were grown on $TiO_2$-terminated (100) STO substrates by PLD in $5*10^{-6}$ Torr oxygen pressure at 710°C, following our previous optimization of SVO growth[6]. A 248 nm wavelength KrF excimer laser was employed with a repetition rate of 1 Hz and an energy density of 1.8 J/cm².

In order to compare the properties of the LAO/SVO heterostructures and the SVO single films, both type of samples were fabricated during the same deposition using motorized shadow masks pressed against the substrate[19]. After the growth of the SVO layer using a first mask creating two separated areas on the substrate surface (step 1 in figure 1), a second mask covering one of the two areas is used to grow LAO on top of the SVO layer (step 2). A thickness of 13 uc of LAO was chosen in order to protect the SVO layer, being thin enough to conduct hard X-ray photoemission spectroscopy measurements. The SVO single layer and the LAO/SVO heterostructure are thus deposited in the same deposition run, allowing to compare the properties directly.

The HAXPES experiments have been conducted at the GALAXIES beamline (Synchrotron SOLEIL, France)[20]. The photon energy was set to 2800 eV and the binding energy scale has been calibrated using the Fermi edge of the sample at 2795.25 eV. The overall resolution was better than 300 meV and all measurements have been done at room temperature. A Shirley background has been subtracted from every core-level spectrum. The inelastic mean free path for V *2p* core-levels (kinetic energy of 2278 eV) is estimated at 3.6 nm by using SESSA software for a probing depth of 10.8 nm at normal emission (90° Take-Off Angle TOA) [21].

SVO film thickness was measured with *ex-situ* X-Ray Reflectivity (XRR) and using the high-angle Laue oscillations. For more details, refer to reference 6. Magneto-transport properties were measured using four in-line aluminum contact pads in a Physical Properties Measurement System (PPMS) in the 2K-300K temperature range.

## 3) Results

### A) Transport measurements

Figure 2a shows the evolution of resistivity as a function of temperature for SVO single films of decreasing thicknesses. A minimum in resistivity is observed for film thickness at 12, 38 and 85 K for the 18, 14 and 8 u. c. samples, respectively. For all thicknesses, the high-temperature metallic part of the ρ(T) curves follows a T² (Fermi liquid) behavior as shown by the dashed line on figure 2a, expressed by $\rho = \rho_0 + A * T^2$, where *A* is a coefficient related to electron-electron scattering and $\rho_0$ is the background contribution due to static disorder[2]. This dependence is characteristic of correlated electrons and expected for SVO thin films. As already observed by Fouchet et al.[6], the values of the coefficient *A*, increasing with decreasing SVO thickness, are similar to the values observed in different SVO studies on bulk and thin film samples[2,22,23], but also to other oxide metal transition materials[24]. $\rho_0$ tends also to increase with decreasing thickness, indicating an increase of the static disorder with the reduction of the thickness. Finally, below 4 u. c. (1.5 nm), the SVO single films show an insulating behavior (Figure 2b).

To compare the SVO single films and the LAO/SVO heterostructure (LAO/SVO), we present on the figure 2d and 2e, the temperature dependence of the resistivity for SVO films of 18 u.c. and 8 u.c. with and without a LAO capping layer, respectively. For both thicknesses, adding a LAO capping layer dramatically changes the conduction behavior. At 300K, the resistivity is lower for both capped films. Furthermore, there is no minimum of resistivity in the ρ(T) curve for the capped 18 u.c. SVO contrary to the single SVO film (see figure 2d). For the 8 u.c. SVO film, adding the LAO layer shifts the minimum in resistivity from 85 to 45 K. Finally, by comparing a SVO layer of 4 u.c. with a LAO/SVO heterostructure, the SVO single film is showing an insulating behavior for all temperatures (see fig. 2b), whereas a minimum of resistivity is observed at 135 K for LAO/SVO film (see fig. 2c). The temperature dependence of the resistivity of the LAO/SVO heterostructures was also compared to SVO single films, and a deviation from the Fermi liquid behavior T² was observed. Such a deviation has also been observed in the $La_{1-x}Sr_xVO_3$ solid solution in the vicinity of the magnetic phase boundary with a $\rho = \rho_0 + B * T^{1.5}$ behavior [25,26]. We have fitted the metallic part of the resistivity with $\rho = \rho_0 + B * T^{1.5} + A * T^2$, for the 18 and 8 u.c. LAO/SVO heterostructures, results are given in table 1. As mentioned before, the SVO single films can be fitted only with a $T^2$ contribution, whereas a clear contribution of the $T^{1.5}$ appears for the LAO/SVO heterostructures, becoming more important with the decrease of the thickness.

| | | $\rho = \rho_0 + B*T^{1.5} + A*T^2$ | | |
|---|---|---|---|---|
| | Thickness | $\rho_0$ (Ω.cm) | $B$ (Ω.cm.K$^{-1.5}$) | $A$ (Ω.cm.K$^{-2}$) |
| SVO | 18 uc | $7.3*10^{-5} \pm 1*10^{-6}$ | 0 | $4.7*10^{-10} \pm 1*10^{-11}$ |
| LAO/SVO | 18 uc | $3.9*10^{-5} \pm 1*10^{-6}$ | $1.0*10^{-9} \pm 1*10^{-10}$ | $2.2*10^{-10} \pm 1*10^{-11}$ |
| SVO | 8 uc | $17.4*10^{-5} \pm 1*10^{-6}$ | 0 | $7.8*10^{-10} \pm 1*10^{-11}$ |
| LAO/SVO | 8 uc | $7.6*10^{-5} \pm 1*10^{-6}$ | $5.7*10^{-9} \pm 1*10^{-10}$ | $1.1*10^{-10} \pm 1*10^{-11}$ |

Table 1: Fit *results of the resistivity data (figure 2) with a $T^{1.5}$ and $T^2$ contribution (see the text for more details).*

Different phenomena can account for the apparition of the $T^{1.5}$ contribution. First, in the case of $La_{1-x}Sr_xVO_3$ solid solution, $T^{1.5}$ behavior at the vicinity of the MIT was interpreted by the presence of antiferromagnetic spin fluctuations. But a $T^{1.5}$ power law has also been reported in the case of oxygen nonstoichiometry in other complex oxides. In $LaNiO_3$, a $T^{1.5}$ dependence arises from localized spin fluctuations induced by oxygen deficiency creating divalent $Ni^{2+}$. As in our case, the $T^{1.5}$ temperature dependence is only observed for the LAO/SVO heterostructures, the presence of the La ions in the vicinity of the interface seems to induce oxygen vacancies in the SVO layer. Furthermore, the fact that the B parameter increases with decreasing thickness underlines the importance of the interfacial region. This scenario is confirmed when comparing the *A* parameter related to the electron–electron scattering, the heterostructure is always showing lower values than the films without LAO. Since *A* is proportional to the square of the effective mass m* and to the Fermi wave-vector $k_F$, a small variation of the shape of the Fermi surface induced by the presence of the LAO layer and/or oxygen vacancies[28] and the related structural distortion may lead to changes of *A*. Moreover, when we compare the single films without LAO layer and with LAO, $\rho_0$ is always higher, meaning a lower static disorder when the film is capped.

Finally, when the thickness of the films were decreased to 3 u.c., the minimum of resistivity is observed at 250 K and for 2 u.c., the heterostructure presents an insulating behavior over the entire temperature range as seen in figure 2c. This critical thickness of 2 – 3 u.c. for the thickness driven MIT is smaller in the LAO/SVO heterostructures compared to the SVO single films.

B) Hard x-ray Photoemission Spectroscopy

We conducted HAXPES experiments on the 3 u.c. SVO single film and LAO/SVO heterostructure. The major advantage of using hard X-ray in photoemission spectroscopy is the ability to reach buried interfaces not accessible in soft x-ray PES. At hν = 2800 eV, we are able to probe the electronic and chemical properties down to 10 nm below the surface.

Figure 3a shows broad-scan surveys (hν = 2800 eV) of the SVO thin film (bottom, gray curve) and LAO/SVO (top, black curve) heterostructure. These spectra show peaks from every expected elements, even from the deeply buried STO substrate (i. e. Ti 2p core-level peak at ca. 450 eV binding energy). Few carbonated and water contamination is visible on the C 1s and O 1s peaks [29]. One of the interests of hard X-ray photoemission spectroscopy is the ability of looking through this contamination without using usual surface cleaning procedure which can alter the thin films properties [30].

Figure 3b and 3c shows the V 2p core-level for LAO/SVO heterostructures and SVO thin films respectively for normal emission. The secondary electron background has been removed using a Shirley background [31]. V 2p spectra are fitted using Voigt functions with full-widths at half maximum (FWHM) of each component of 1.3 eV for V $2p_{3/2}$ and 2.4 eV for V $2p_{1/2}$. This difference is likely due to the Coster-Kroning effect, a phenomenon well known for instance in Ti 2p core-level[32]. Both spin-orbit-split levels (splitting of 7.2 eV) show three sub-components for both samples. We label these components A, B and C for decreasing binding energies (see Table 2). On one hand, these three components are usually allocated to Vanadium in a 5+/4+/3+ oxidation state (with decreasing binding energy) [33,34]. On the other hand, it has been recently shown that even optimally oxidized SVO still shows a $V^{5+}$ and $V^{3+}$ component in addition to the expected $V^{4+}$ component [35]. According to Lin et al. who show such spectra on $SrVO_3$, the three Vanadium core-level components can be associated with different d0/d1/d2 final states (for decreasing binding energy) associated with charge fluctuations of metallic $SrVO_3$ [35]. The possible coexistence of different oxidation (initial state phenomenon) and final-state effects in such samples makes any quantitative conclusions on the $V^{4+}/V^{3+}$ ratio quite complicated. However, a purely qualitative study of the HAXPES results is sufficient to better understand the transport behavior of our thin film.

When going from LAO/SVO to SVO, the spectral weight of peak A increases strongly (see Table 2). This cannot be only due to a final state effect, which should not change with the added overlayer. This increase indicates rather that a significant part of vanadium at the interface of the heterostructures or at the surface of the single films is in a 5+ oxidation state. Recently, we have shown that an insulating $Sr_3V_2O_8$ phase can form at the surface of SVO when the surface is in contact with oxygen at high temperature[28], as is the case for instance after the film growth in a PLD chamber. In this phase, vanadium in a 5+ oxidation state is expected. Therefore, we can conclude that (1) SVO has tendency to oxidize as expected from reference 28 and (2) capping SVO with an ultrathin LAO layer prevents in a certain measure this chemical reconstruction. As shown in Supplementary Information, complementary experiments with grazing electron emission confirm our interpretation. In LAO/SVO, the $V^{5+}$-related component increases relatively to the two other components for interface sensitive measurements, suggesting an interface layer with a different oxidation degree than that expected for SVO. On the contrary, there is no clear trend for the single SVO films where the $V^{5+}$ spectral weight is similar for both bulk and surface sensitive measurements. Therefore, in the single SVO films, the $V^{5+}$-phase is extended over the full 3 u.c. thickness. This confirms that LAO acts as a protective layer for the SVO thin film, though the interface itself still is overoxidized. We interpret this gradient as inherent to the deposition process itself, where the surface of the SVO film begins to oxidize at high temperature and the deposition of LAO layer stops the oxidation process, explaining the observed difference in the transport properties of capped SVO compared to single films for the same SVO thickness.

By simply adding a very thin, transparent layer of LAO, we achieve a metal-insulator transition in SVO for lower thicknesses paving the way to full-oxide conducting, transparent electrodes.

| Component | A - $V_{5+}$ $d^0$ | B - $V_{4+}$ $d^1$ | C – $V_{3+}$ - $d^2$ |
|---|---|---|---|
| Component binding energy (eV) | -517.725 | -516.35 | -514.95 |
| SVO area ratio | 49 | 26 | 25 |
| LAO/SVO area ratio | 39 | 43 | 18 |

Table 2 *A summary of the fits obtained from our hard X-ray photoemission spectroscopy experiments (see Fig.3).*

As magnetoresistance (MR) gives valuable information about the physics and also information about defects in vanadate films [36–38], we compare the MR of the LAO/SVO film with SVO film. As seen for SVO films of 3 nm[6], positive magnetoresistance is observed and a very small negative magnetoresistance develop only at 2K for the 1.5 nm SVO film[6]. This positive magnetoresistance of the SVO films has a H² dependence and was attributed to the Lorentz force. Furthermore at low magnetic field, a quick rise of the positive magnetoresistance was also observed for the thinnest films and was attributed to weak antilocalization at low temperature.

Surprisingly, for a LAO/SVO heterostructure of 3-4 u.c., (see inset fig. 4), a negative magnetoresistance of -1.2% is observed at 2K. Two possibilities can explain this negative magnetoresistance: first, a spin dependent scattering of the conduction electrons by localized magnetic moments, or a weak localization due to disorder. In any case, this magnetoresistance can be attributed to oxygen vacancies present in the LAO/SVO heterostructure.

In order to further investigate the role of the oxygen vacancies, we annealed the LAO/SVO heterostructure in an oxygen pressure of $PO_2=1.5*10^{-6}$ Torr[28] at 700°C. Interestingly, after annealing, the resistivity versus temperature is almost the same as that before annealing (see figure 4), whereas the magnetoresistance becomes positive, comparable to what was observed without annealing for single SVO films (see inset fig 4). This is also showing that growth of LAO on top of SVO induces oxygen vacancies in SVO, as was observed for LAO/STO interface[1,39].

In this paper, we show that the MIT in ultra-thin films of SVO can be controlled by interface engineering, embedding SVO between the two insulators STO and LAO. Such LAO/SVO heterostructures have shown interesting features compared to single SVO films on an STO substrate, as for example a decrease of the resistivity and a lower critical thickness for the MIT.

HAXPES measurements have shown the importance of different oxidation states on the vanadate films. The LAO layer allows to avoid the oxidation of vanadium from $V^{4+}$ to $V^{5+}$ in the entire SVO..

The magnetoresistance brings also new properties at low temperature with a negative magnetoresistance which can be attributed to magnetic scattering or disorder due to the presence of oxygen vacancies. In the case of vanadates, chemical reconstruction provides a dead layer at the surface of the thin film which can drastically tune the electronic properties when the thickness of the film becomes comparable to the thicknes of the perturbed surface. Thus, it is important to protect the surface of SVO films with a capping layer, leading to more stable ultrathin films of SVO at ambient atmosphere.

## 5) Summary

The present study highlights that using a LAO capping layer allows the fabrication of oxide, transparent electrodes with reduced thickness for future oxide electronics application. Such heterostructures provide a possibility to tune the properties at the interface as for LAO-STO system. These interfaces are a complex playground where the electronic properties can be tuned with different mechanism such as oxygen stoichiometry, polar discontinuity, interdiffusion, strain effects. This study is showing for the vanadate films also the importance of chemical reconstruction which is a way to control of the electronic properties and stabilize a metallic SVO film for the thickness limit of the MIT to 2 -3 u.c. at ambient atmosphere.

**Supplementary Material:**

Comparison of Normal and Grazing emission of SVO and LAO-SVO heterostructure.

**Aknowledgments:**

This work was supported by Ile de France region for magneto-transport measurements ("NOVATECS" C'Nano IdF project n°IF-08-1453/R). This work is supported by a public grant from the "Laboratoire d'Excellence Physics Atom Light Mater" (LabEx PALM) overseen by the French National Research Agency (ANR) as part of the "Investissements d'Avenir" program (reference: ANR-10-LABX-0039). Preliminary experimental support by W. Prellier, U. Lüders, A. David and H.

Rotella is acknowledged. We also acknowledge SOLEIL for provision of synchrotron radiation facilities.

Figure 1: Illustration of the deposition process for the SVO film and the LAO/SVO heterostructure.

Figure 2: (a) Resistivity versus temperature for a 4 u.c. thick single SVO film. Inset: Resistivity versus $T^2$ for different SVO film thicknesses . (b) Resistivity versus temperature for different LAO/SVO heterostructures (b). Comparison of SVO and LAO/SVO transport measurement for 18 u.c. (c) and 8 u.c. (d). The red dash curve is the fit obtained of the metallic part with $\rho = \rho_0 + B*T^{1.5} + A*T^2$ for the films of 8 and 18 u. c. (see table 1).

Figure 3: Broad-scan surveys (hν = 2800 eV) of the SVO 3. u. c. (bottom, gray curve) and LAO/SVO 3 u. c. (top, black curve) thin films (a) this latter curve was shifted vertically for convenience. V 2p core-level for LAO/SVO (b) and SVO thin film (c) respectively for normal emission. The green, blue and red contributions are referred as A ($V^{5+}$) B ($V^{4+}$) C ($V^{3+}$) of table 2 respectively.

Figure 4: Resistivity versus temperature of 3-4 u.c. SVO thick film in the heterostructure LAO/SVO before and after annealing at 700°C with $PO_2 = 1.5*10^{-6}$ Torr. In inset the Magnetoresistance at 2K as deposited and after annealing.

Table 1 *A summary of the fits obtained from our transport experiments.*

Table 2 *A summary of the fits obtained from our* (HAXPES) *experiments.*

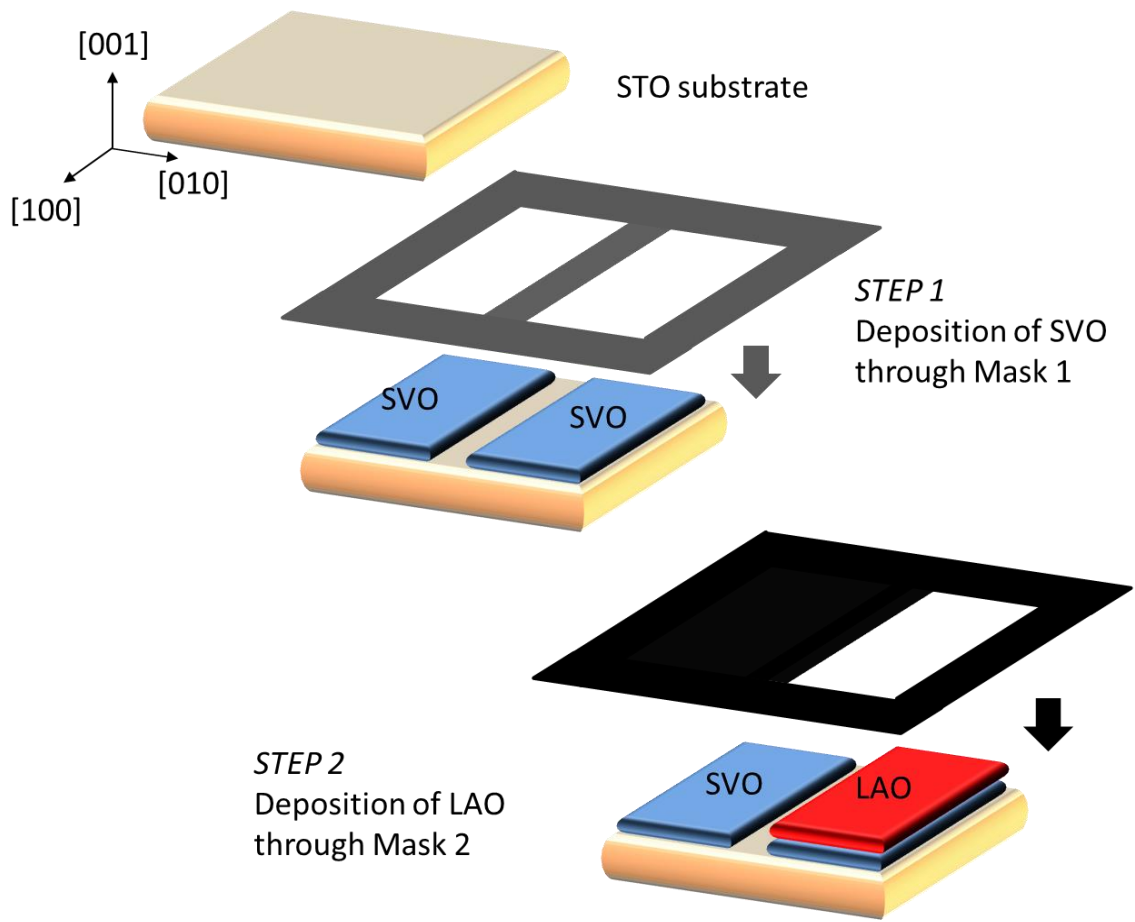

Figure 1

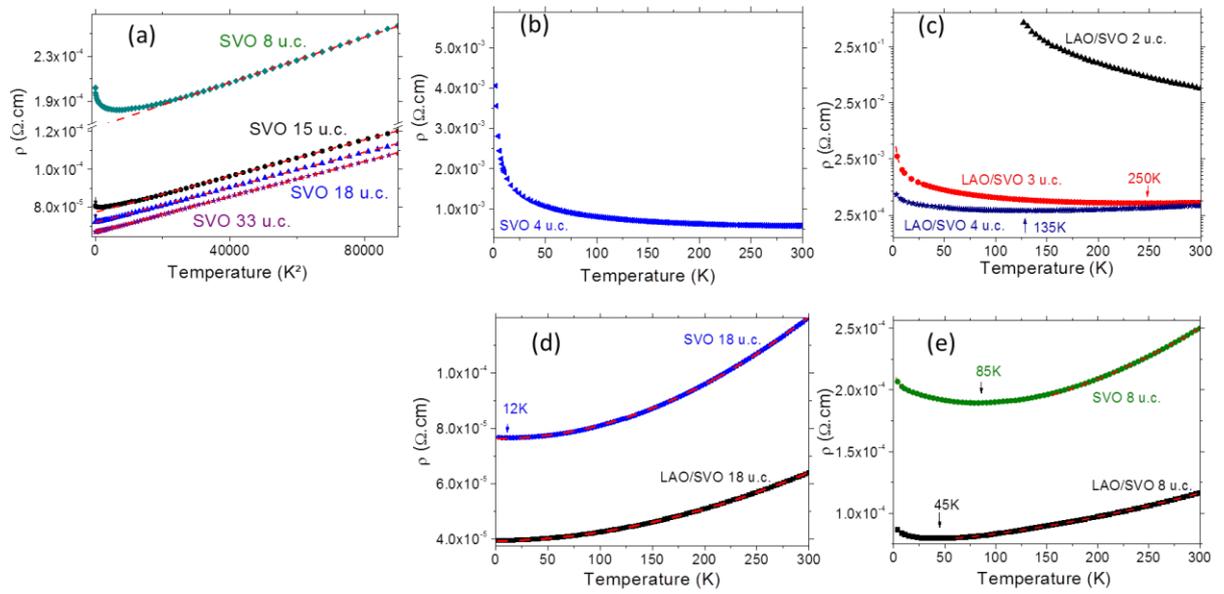

Figure 2

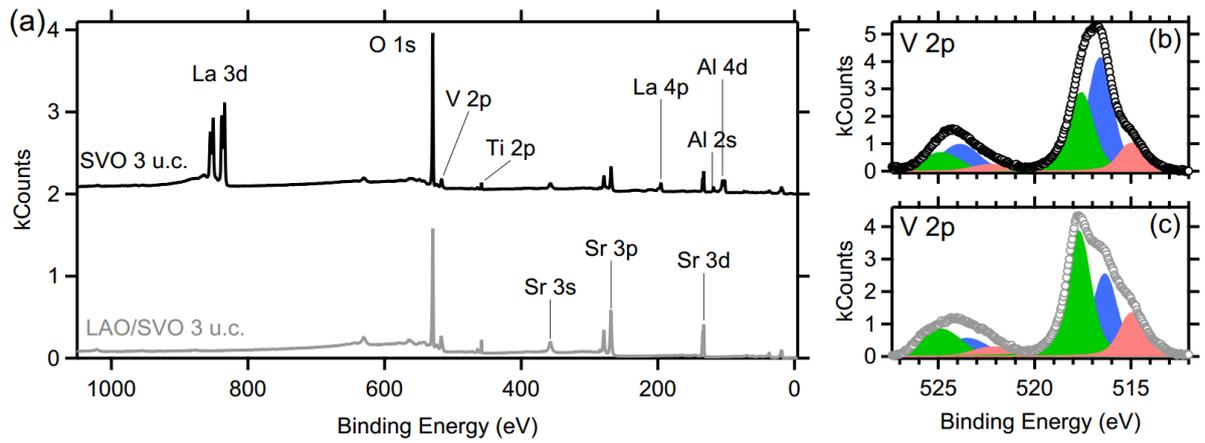

Figure 3

LAO/SVO Film 3 - 4 u.c.

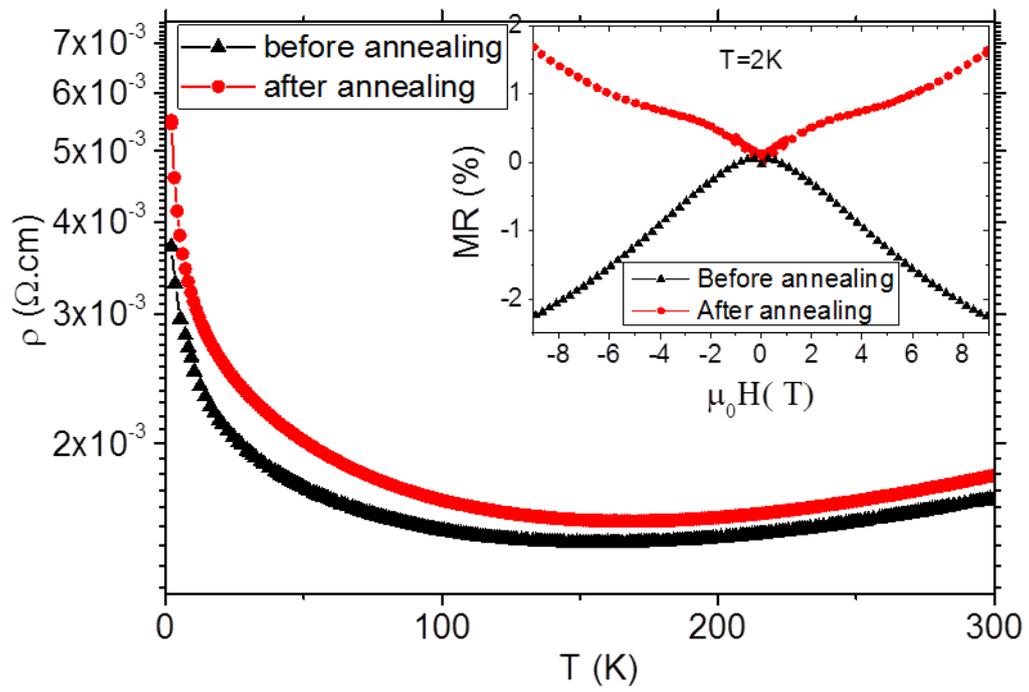

Figure 4

# Supporting information of

**1/ Comparison of Normal and Grazing emission of SVO and LAO-SVO heterostructure:**

Additional angle-dependent X-ray photoemission spectroscopy (HAXPES) experiments have been conducted at GALAXIES beamline (Synchrotron SOLEIL, France)[1]. The photon energy was set at 2800 eV and the binding energy scale has been calibrated using the Fermi edge of the sample at 2795.25 eV. The overall resolution was better than 300 meV and all measurements have been done at room temperature. A Shirley background has been subtracted from every core-level spectrum. The inelastic mean free path for V *2p* core-levels (kinetic energy of 2278 eV) is estimated at 3.6 nm for a probing depth of 10.8 nm at normal emission (90° Take-Off Angle TOA) and 7.8 nm at grazing emission 35° (TOA) (calculated via SESSA software)[2].

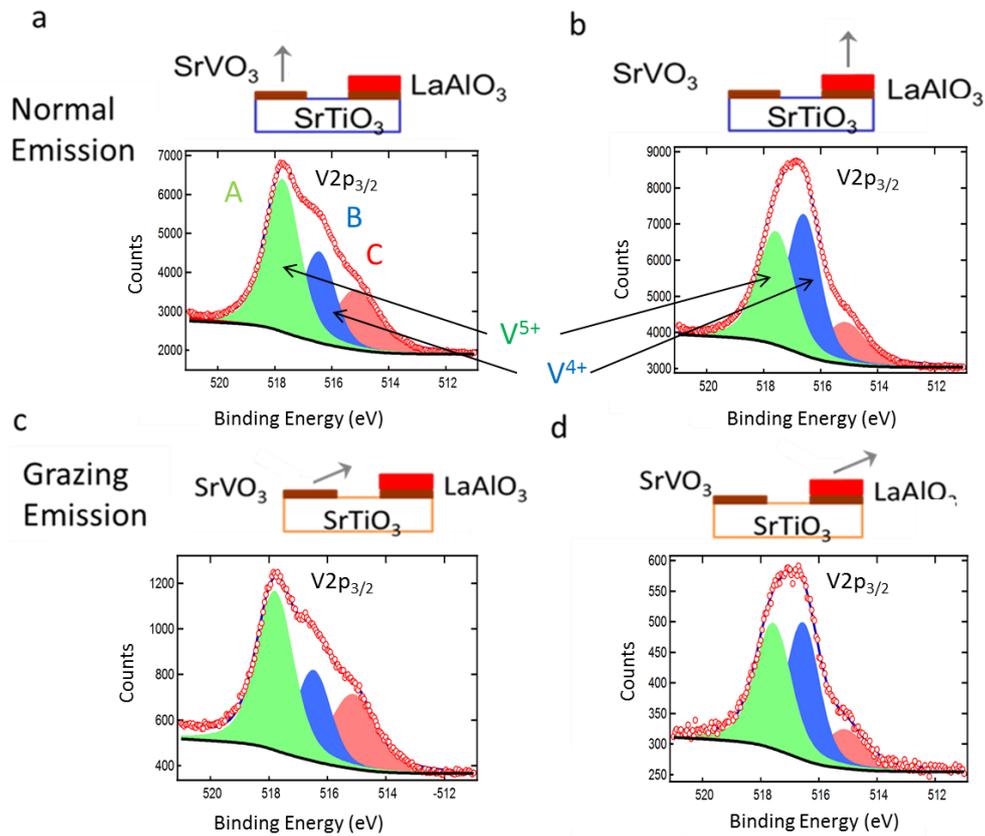

Figure 1-SI: Normal Emission (NE) of V 2p core-level for SVO film (a) and LAO/SVO (b) and Grazing Emission (GE) for SVO (c) and LAO-SVO (d). The green, blue and red contributions are referred as A ($V^{5+}$) B ($V^{4+}$) C ($V^{3+}$) of table 1-SI respectively.

| Component | A - $V_{5+}$ $d^0$ | B - $V_{4+}$ $d^1$ | C – $V_{3+}$ - $d^2$ |
|---|---|---|---|
| Component binding energy (eV) | -517.725 | -516.35 | -514.95 |
| SVO area ratio NE | 49 | 26 | 25 |
| SVO area ratio GE | 48 | 24 | 28 |
| LAO/SVO area ratio NE | 39 | 43 | 18 |
| LAO/SVO area ratio GE | 43 | 41 | 16 |

Table 1-SI *summary of the fits obtained from our HAXPES experiments in Normal Emission and Grazing Emission from fig. 1-SI.*

On LAO/SVO, the $V^{5+}$-related component increases for interface sensitive measure: there is an interface layer with a different oxidation degree. On the contrary, there is no clear trend for plain SVO: $V^{5+}$ is similar for the two angles. This shows the SVO layer $V^{5+}$-phase is extended over the full (extremely low) thickness of the SVO after air exposure. It's not the case of LAO-protected SVO which only shows an interface-related component.

Thus the grazing emission experiment is showing that LAO prevent the formation of higher oxidation state in the first few layers of SVO but still a stiff gradient is observed at the interface. We interpret this gradient as inherent to the deposition process itself, where the surface of the SVO film begins to oxidize at high temperature and deposition of LAO layer stops the oxidation process.